\documentclass[aps,nofootinbib]{revtex4}
\usepackage{epsfig,bm,latexsym,amssymb}
\usepackage{theorem}

\def\B.#1{{\bm{#1}}}
\def\C.#1{{\cal #1}}
\theorembodyfont{\upshape} 
\newtheorem{proposition}{Proposition}[section]

\begin{document}
%%%%%%%%%%%%%%%%%%%%%%%%%%%%%%%%%%%%%%%%%%
\title{Algebraic decay in hierarchical graphs} 
\author{Felipe Barra}
\altaffiliation[Permanent address: ]{Dept. F\'{\i}sica, Facultad de ciencias 
F\'{\i}sicas y Matem'{a}ticas universidad de Chile, casilla 487-3 Santiago,
Chile.} 
\affiliation{Department of~~Chemical Physics, The
Weizmann Institute of Science, Rehovot 76100, Israel}
\author{Thomas Gilbert}
\affiliation{Department of~~Chemical Physics, The
Weizmann Institute of Science, Rehovot 76100, Israel}

%\date{version of \today}
%%%%%%%%%%%%%%%%%%%%%%%%%%%%%%%%%%%%%%%%%%%%%%%%%%%%%%%%%%%%%%%%%%%%%
\begin{abstract}
We study the algebraic decay of the survival probability in open
hierarchical graphs. We present a model of a persistent random walk on a 
hierarchical
graph and study the spectral properties of the Frobenius-Perron operator.
Using a perturbative scheme, we derive the exponent of the classical 
algebraic decay in
terms of two parameters of the model. One parameter defines the geometrical 
relation between the length scales on the graph, and the other relates to
the probabilities for the random walker to go from one level of the
hierarchy to another. 
The scattering resonances of the corresponding hierarchical
quantum graphs are also studied. The width distribution shows
the scaling behavior $P(\Gamma) \sim 1/\Gamma$.
\end{abstract}

\maketitle
%%%%%%%%%%%%%%%%%%%%%%%%%%%%%%%%%%%%%%%%%%%%%%%%%%%%%%%%%%%%%%%%%%%%%

%%%%%%%%%%%%%%%%%%%%%%%%%%%%%%%%%%%%%%%%%%%%%%%%%%%%%%%%%%%%%%%%%%%%%
\section{introduction}
%%%%%%%%%%%%%%%%%%%%%%%%%%%%%%%%%%%%%%%%%%%%%%%%%%%%%%%%%%%%%%%%%%%%%
Typical Hamiltonian systems are non-integrable and have a mixed phase
space, where regions of regular and chaotic motions coexist. The chaotic 
dynamics of mixed systems is clearly different from the fully chaotic case. 
This is manifest in the behavior of the survival probability in open
systems. 

Assume we have an infinite hierarchy of Kolmogorov-Arnold-Moser (KAM) small 
islands interspersed in a connected chaotic region, and suppose we draw a
boundary at a given level of this hierarchy, such that the particles
leaving this boundary are lost.
Consider a large initial number $N_0$ of randomly
chosen (with respect to a given probability distribution) initial
conditions in the chaotic region and let them evolve by
the dynamics up to some time $t$. The survival probability $P(t)$ is the
ratio $N(t)/N_0$ in the limit of large $N_0$, where $N(t)$
is the number of particles remaining within
the boundary at time $t$. In the typical case this
probability is believed to decay algebraically,
\begin{equation}
P(t) \sim t^{-\delta} \label{alg-decay}\ .
\end{equation}
It has been argued that the algebraic decay is due to the hierarchical 
structure of phase space \cite{84CS,85HCM,84MMP,87MMP,85MO,86MO}. However,
despite significant efforts, the mathematical understanding of the behavior 
described by Eq. (\ref{alg-decay}) is rather poor \cite{92M}.
Much of our current knowledge of this problem is based on 
the self-similar Markov chain model \cite{85HCM,85MO}, which provides an 
expression for the exponent $\delta$ in terms of the parameters of the model.
Yet a precise and simple understanding of
the mechanism based on dynamical properties is lacking. 

In fully chaotic open systems, the survival probability decays exponentially,
\begin{equation}
P(t) \sim e^{-\gamma t}\ . \label{exp-decay}
\end{equation}
This case is well understood. The evolution operator of the probability
densities, the Frobenius-Perron operator, admits a spectral decomposition 
in terms of Pollicott-Ruelle resonances\footnote{Note that the use of the term
  resonances here is restricted to the logarithms of eigenvalues of the 
  Frobenius-Perron
  operator, as opposed to its use in the KAM theory, e.~g. as in
  \cite{87MMP}.} 
\cite{PoRu} which characterize the
relaxation properties.  In particular, for open systems, the leading resonance
is identified as the escape rate $\gamma$ in Eq. (\ref{exp-decay}), and 
describes the slowest relaxation mode of the probability distributions. 
We point out that in closed systems an equilibrium state exists (the
leading resonance is equal to zero), and one
can study the relaxation to 
this equilibrium state by considering the next leading resonance. In
contrast, for open systems the final state does not exist due to the escape,
the rate of which is characterized by the leading resonance of the
Frobenius-Perron operator. We refer to \cite{Gasp98} for more details
concerning the connection between open and closed systems.

The escape rate can also 
be interpreted as a macroscopic quantity resulting e.~g. from a diffusion
process described by a Fokker-Planck equation for the macroscopic density
of particles. This connection between microscopic dynamics and macroscopic
processes, known as the escape rate formalism
\cite{Dorf99,Gasp98,GN90,GD95,DG95}, 
yields expressions of the
transport coefficients, e.~g. the diffusion coefficient, in terms of
the dynamical quantities. The existence of this
connection relies heavily on the hyperbolic properties of the system, 
i.~e. (i) (almost) every point in phase space is assumed to be of saddle type, 
and (ii), for the open boundaries, the repeller is fractal.

The absence of an exponential decay rate of the survival probability for a
typical system
Eq. (\ref{alg-decay}) is associated to anomalous transport, e.~g. in a
diffusive process the mean
square displacement grows with a power of $t$ not equal to 1.
With this respect, the connections between 
macroscopic phenomena and microscopic dynamics are far less understood for
typical systems than they are in the fully chaotic case. 

Attempts to describe the relaxation properties of systems with mixed phase
space in terms of spectral properties of the Frobenius-Perron operator have
introduced regularization procedures which amount to truncating the
Frobenius-Perron operator in a finite matrix representation
\cite{WHS2000}. An alternative approach \cite{KFA2000} considers the
presence of a vanishing noise and yields finite values of the leading
relaxation rates. In both these approaches, it is worthwhile stressing that 
the relaxation rates are the analogues of the leading Pollicott-Ruelle
resonances mentioned above in reference to relaxation in fully chaotic 
systems. 

Our purpose in this paper is to understand what properties of the
Pollicott-Ruelle spectrum characterize the algebraic as opposed to
exponential decay of the survival probability. We will do so by considering 
a model whose finite approximations are fully chaotic, but which displays
algebraic decay of the survival probability as a limiting property.

For the purpose of this endeavor, we propose to investigate the decay 
properties of an open one-dimensional hierarchical graph, whose survival
probability turns out to decay algebraically. A graph is a
collection of bonds on which a classical particle has a uniform
one-dimensional motion. The bonds are interconnected by vertices where
neighboring bonds meet. At the vertices, the particle undergoes a 
conservative collision process with the result that its velocity may change 
direction. In practice this collision process is determined by a random
process where outputs are assigned fixed transition probabilities in terms
of the inputs.
A hierarchical graph is one where the lengths of the bonds and the 
transition probabilities obey scaling laws \cite{ketzmerick}.

The specific model we propose to study is based on a one-dimensional
Lorentz lattice gas \cite{vV90,DE95}, the difference being that ours
is a continuous time process where the
separation between scatterers will be taken to satisfy a scaling law. The
scattering probabilities depend on the direction of the particle, in
analogy to the Lorentz lattice gas, with the further property that these
probabilities change according to the index of the scatterer.
Due to its connection to persistent random walks, we propose to refer to
our model as a {\em persistent hierarchical graph}.
In such a system, the evolution operator for phase space densities, the
Frobenius-Perron operator, can be written explicitly and its spectral
decomposition expressed in terms of the Pollicott-Ruelle resonances 
$s_j$ \cite{Gas96}. This in turn yields the expression for the 
survival probability~:
\begin{equation}
P(t)=\sum_{j=0}^{\infty} A_j e^{s_jt} \ , 
\label{spectral2}
\end{equation} 
where the amplitudes $A_j$ can be expressed in terms of the eigenstates
associated to the corresponding resonances.
The Pollicott-Ruelle resonance spectrum is located in the lower half-plane, 
$\mathrm{Re} \, s_j<0$. We note that the Perron-Frobenius operator is
here defined on a rigged Hilbert space, whose dimension is infinite
\cite{Gasp98}. 

It is a general property that finite open graphs are fully chaotic
systems. Indeed there is a gap empty of resonances, i.~e. the closest 
resonance to the imaginary axis is real and isolated, so that it dominates
the sum in Eq. (\ref{spectral2}). This resonance is the escape rate. Thus
in this case the survival probability decays exponentially as in 
Eq. (\ref{exp-decay}).

On the contrary, in the semi-infinite open hierarchical graph, because the 
lengths of the bonds and the transition probabilities scale in terms of some 
parameters (see Sec. \ref{sec.model}), we will see that the decay of
the survival probability is algebraic as in Eq. (\ref{alg-decay}).
Indeed the power law behavior can emerge
in the limit of infinite graphs from the expression (\ref{spectral2}),
because there is an accumulation of resonances going to zero and distributed
with a particular density.
In fact, if the amplitudes and the decay rates satisfy
$A_j=a \alpha^j$ and $s_j=-b \beta^j$ with $\alpha$, $\beta$, 
$a$ and $b$ some real functions of the parameters of the model 
($0<\alpha$, $\beta<1$), 
then, evaluating the sum in Eq. (\ref{spectral2}) by the steepest decent 
method we get the power law decay of Eq. (\ref{alg-decay}) with
\begin{equation}
\left.\begin{array}{l}
A_j=a \alpha^j\\
s_j=-b \beta^j
\end{array}
\right\}\Rightarrow
\delta = \frac{\ln \alpha}{\ln \beta}
\label{delta}
\end{equation}
We will show in Sec. \ref{sec.pert.th} that indeed these scaling behaviors for
the spectrum $s_j$ and for the amplitudes $A_j$ hold for persistent 
hierarchical graphs. We further point out a connection between the parameters
$\alpha$ and $\beta$ and the scaling parameters of dynamical traps
\cite{98Z,99Z}~: $\alpha$ is the spatial scaling parameter and $1/\beta$ the
temporal one. We will comeback to this in the conclusions.

As already mentioned, every finite size approximation of a persistent
hierarchical graph is a fully chaotic system. This provides means of
making further comparisons between typical and fully chaotic systems.
Dynamical quantities, such as the topological pressure (henceforth referred 
to as free energy) will be considered. 

Recent studies consider the  question how the hierarchical
structure and the dynamics of a typical Hamiltonian system shows
up in quantum properties. 
It was shown by a semi-classical argument \cite{ketzmerick0} that 
Eq. (\ref{alg-decay}) leads to fractal conductance 
fluctuations on an energy scale larger than the mean level spacing.
This fractal conductance 
fluctuations are decorated with peaks corresponding to isolated resonances
at a small energy scale \cite{ketzmerick}, which
are associated to ``hierarchical'' states and are distributed
according to $p(\Gamma) \sim 1/\Gamma$, with $\Gamma$ the width
of the scattering resonance in the wavenumber plane, to be defined in
Sec. \ref{sec.qres}. 

Quantum properties of graphs are not without their own interest.
Spectral properties of closed quantum graphs have been considered in 
\cite{Smilansky} where quantum graphs were introduced for the first 
time as a model for quantum chaos. Other spectral properties were considered 
in \cite{Smilansky2,Keating,art3}. 
Dynamics \cite{BGsub}, scattering \cite{Smilansky3} and localization in 
infinite disordered graphs \cite{Schanz} have also been considered.
On the other hand the classical dynamics has been studied in detail 
in \cite{BG2001}. 
Here we will also study properties of the scattering resonances
in a quantum realization of the persistent hierarchical graph. 

The plan of the paper is as follows. In Sec. \ref{sec.gengraph}
we discuss general properties of classical and quantum graphs and introduce
the persistent hierarchical graphs in some detail. 
The survival probability is defined in Sec. \ref{sec.stay-prob}, where we
give its expression in terms of the spectral decomposition of the 
evolution operator.  Section \ref{sec.pert.th} presents the calculation of 
the spectrum of the persistent hierarchical graph.
Some properties of the free energy are studied in Sec. \ref{sec.pres}.
In Sec. \ref{sec.qres} we turn to the quantum description of persistent
hierarchical graphs and,
in particular, analyze the spectrum of scattering resonances. Finally
conclusions and perspectives are drawn in Sec. \ref{sec.conc}.

%%%%%%%%%%%%%%%%%%%%%%%%%%%%%%%%%%%%%%%%%%%%%%%%%%%%%%%%%%%%%%%%%%%%%
\section{Hierarchical graphs and our model}
\label{sec.gengraph}
%%%%%%%%%%%%%%%%%%%%%%%%%%%%%%%%%%%%%%%%%%%%%%%%%%%%%%%%%%%%%%%%%%%%%
\subsection{General Survey} 
\label{sec.graph}
%%%%%%%%%%%%%%%%%%%%%%%%%%%%%%%%%%%%%%%%%%%%%%%%%%%%%%%%%%%%%%%%%%%%%

A graph is a collection of $B$ one-dimensional bonds, connected by
vertices, where a particle moves freely. 
The position of the particle on the graph is described by a coordinate $x_b$.
The index $b$ refers to a particular bond and $x_b$ to
the position on that bond, with $0<x_b<l_b$, where $l_b$ denotes the length 
of the corresponding bond.
Here we consider oriented graphs where bonds have directions. 
Thus to each ``physical'' bond corresponds two
oriented bonds, and therefore the number of oriented bonds is $2B$.

In a quantum graph the dynamics of the particle on a bond
is governed by the free Schr\"{o}dinger equation.
When the particle arrives to a vertex, a scattering process determines the 
{\em probability amplitude} $\sigma_{bb'}$ for being reflected or
transmitted to the other connected bonds.

These systems admit a classical limit which corresponds to a
particle moving at constant velocity on the bonds and undergoing a
conservative scattering process at the vertices (the classical limit of the 
quantum scattering process), determined by the {\em probabilities} 
$P_{bb'}=|\sigma_{bb'}|^2$
of being reflected or transmitted to other bonds. 
The classical dynamics is Markovian, i.e. there is no memory effect.

Open or {\em scattering} graphs, have some infinite leads $c$, from where
a particle can escape and never return.

Hierarchical graphs are a particular class of graphs consisting of a
self-similar collection of unit cells, which are topologically identical and
whose characteristic lengths follow a given scaling law. In the classical
case, the transition probabilities are taken to satisfy a scaling property, 
such as in a continuous Markov chain \cite{85HCM}. A possible example is
given by a random walk on a fractal support, such as the Sierpinsky gasket, 
where scattering probabilities change according to the level (with respect
to the fractal structure) of the vertices. A simple such example is the
persistent hierarchical graph we introduce below. Other examples are the 
quantum version of the chain model \cite{ketzmerick}, or the Cayley tree for
quantum conduction \cite{shapiro,berko,naim}. 

%%%%%%%%%%%%%%%%%%%%%%%%%%%%%%%%%%%%%%%%%%%%%%%%%%%%%%%%%%%%%%%%%%%%%
\subsection{Persistent Hierarchical Graph}
\label{sec.model}
%%%%%%%%%%%%%%%%%%%%%%%%%%%%%%%%%%%%%%%%%%%%%%%%%%%%%%%%%%%%%%%%%%%%%

The hierarchical graph that we consider is a semi-infinite one-dimensional
lattice where the lengths of the bonds decay exponentially 
with the bond index. On this lattice, a random walker moves on the bonds
with constant speed, so that the time between collisions becomes exponentially 
shorter as the walker moves deeper into the lattice. Moreover, at each vertex,
the walker undergoes a random collision which reverses its direction 
%and keep  it on the same bond 
with some probability $q_n$, which depends on the index $n$ 
of the scattering vertex, or keeps the direction of the walker unchanged 
with probability $p_n$.
% transmits it to the next bond . 
We will label by $n$ the (non-directed) bond between vertices $n$ and
$n+1$. A  directed bond $b$ is either $(n,+)$ or $(n,-)$. 
In terms of those, the following transitions $P_{bb'}$ of going from $b'$ to
$b$ are possible~:
\begin{eqnarray}
b'&=&(n,+)\longrightarrow\left\{\begin{array}{l@{\quad{\rm with\,probability}\quad}l}
b=(n+1,+),&P_{(n+1,+),(n,+)}=p_{n+1}\ ,\\
b=(n,-),&P_{(n,-),(n,+)}=q_{n+1}\ .
\end{array}
\right. \label{proba1}\\
b'&=&(n,-)\longrightarrow\left\{\begin{array}{l@{\quad{\rm with\,probability}\quad}l}
b=(n-1,-),&P_{(n-1,-),(n,-)}=p_{n}\ ,\\
b=(n,+),&P_{(n,+),(n,-)}=q_{n}\ .
\end{array}
\right. \label{proba2}
\end{eqnarray}
All the other probabilities are zero.

The probabilities $p_n$ and $q_n$ are chosen so as to favor backscattering
of particles as they move deeper into the lattice,
\begin{equation}
\begin{array}{l}
p_n = p_0 \varepsilon^n\ ,\\
q_n = 1 - p_n\ ,
\end{array}
\label{pn}
\end{equation}
where $0<\varepsilon<1$ is a fixed parameter.

\begin{figure}
\epsfxsize=12truecm
\epsfysize=6truecm
\epsfbox{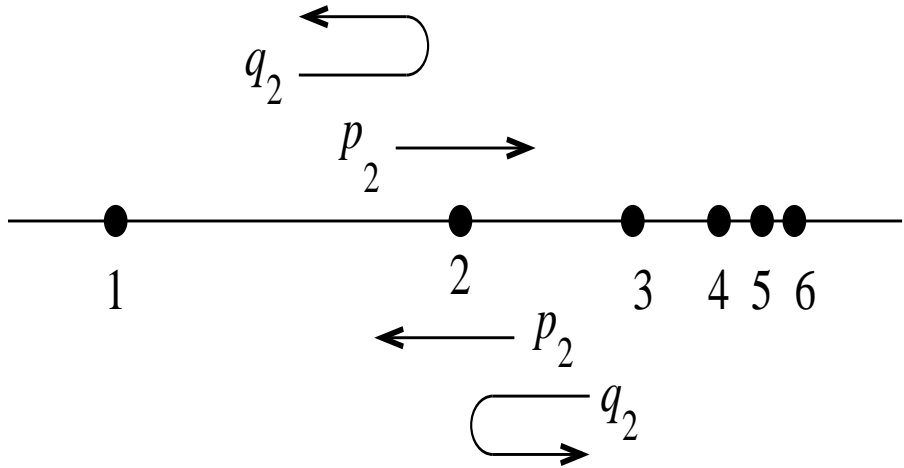}
\caption{Schematic illustration of the possible transitions and their
  probabilities for a particle colliding with a scatterer. The
  parameter $\mu$ is here taken to be $1/2$. There are a total of $5$ bonds 
  and $2$ scattering leads in this example.
}
\label{fig.graph}
\end{figure}

We study this infinite hierarchical graph by considering finite approximations 
made of $N+1$ vertices with open boundary conditions
at both ends. That is, infinite leads are connected to vertices $1$ and
$N+1$ respectively from the left and right.
%That is at the vertex $1$ there is a scattering lead connected
%to the bond 1 and the vertex $N+1$ connect the $N$ bond with another scattering lead.
For the length of the bonds $l_n$ we assume
\begin{equation}
\begin{array}{ll}
l_n = l_0 \mu^n\ ,&1\le n \le N\ ,
\end{array} 
\label{links}
\end{equation}
where $l_0$ is an arbitrary length scale, which we will set to one, and 
$\mu$, $0<\mu<1$, is a (dimensionless) fixed parameter. We note that the
total length
of the lattice is $L_{N}= l_0 \mu(1 - \mu^{N+1})/(1 - \mu)$, which is bounded by
$l_0 \mu/(1-\mu).$ A schematic representation of the system is presented in
Fig. \ref{fig.graph}. 
%%%%%%%%%%%%%%%%%%%%%%%%%%%%%%%%%%%%%%%%%%%%%%%%%%%%%%%%%%%%%%%%%%%%%%%%%%%
%%%%%%%%%%%%%%%%%%%%%%%%%%%%%%%%%%%%%%%%%%%%%%%%%%%%%%%%%%%%%%%%%%%%%%%%%%%

\begin{figure}
\epsfxsize=12truecm
\epsfysize=8truecm
\epsfbox{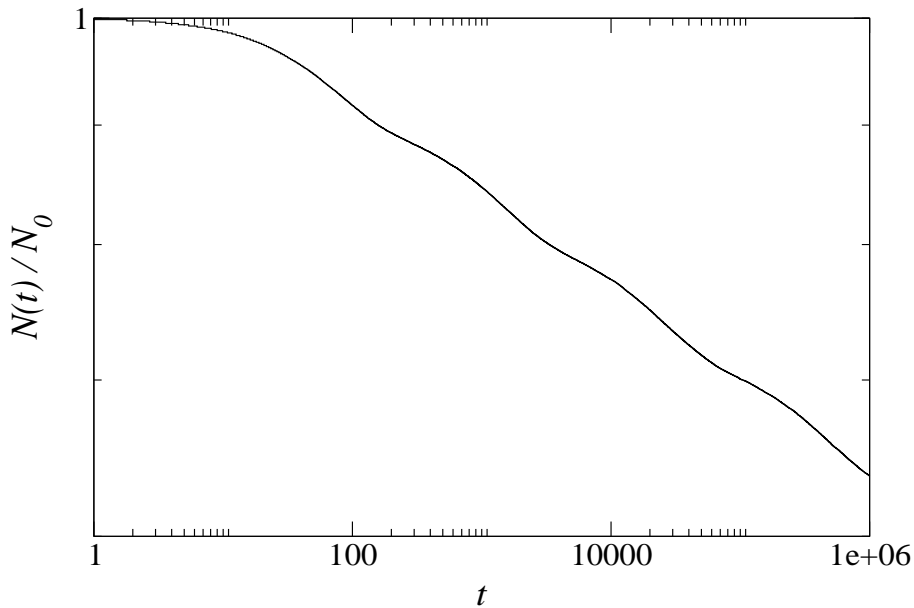}
\vskip 2mm
\caption{Fraction $N(t)/N_0$ of particles remaining in the 
system after time $t$. The system has $25$ bonds and $N_0=100,000$ 
particles
are initially distributed at random positions ({\em i. e.} evenly with
respect to the position on the line). The parameter values are
$\varepsilon=1/20$ and $\mu=9/10$. Each particle is run for a maximal time
$10^6$ and escape times are recorded, which yields the fraction of
particles of surviving particles vs. time.
}
\label{fig.decay}
\end{figure}

\begin{figure}
\vskip 2mm
\epsfxsize=12truecm
\epsfysize=8truecm
\epsfbox{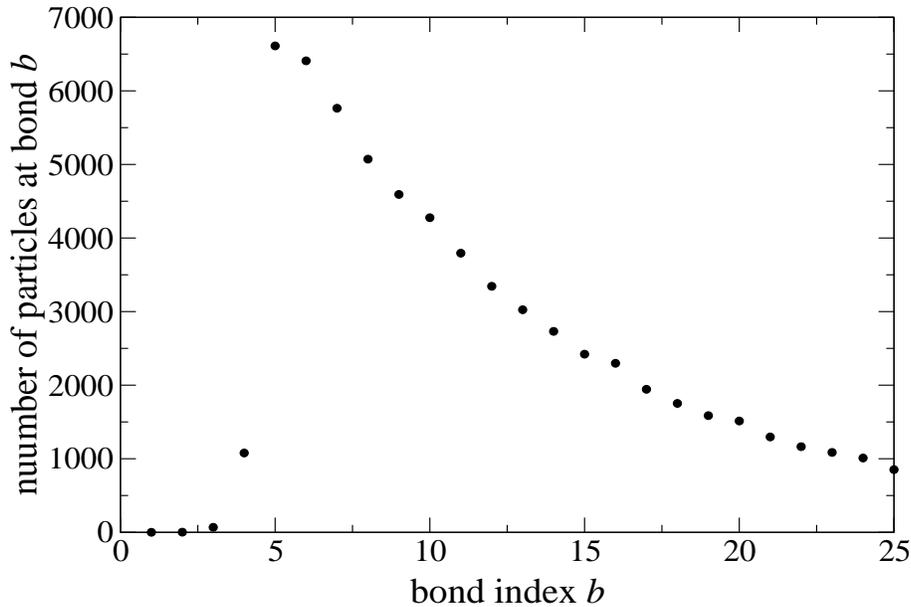}
\vskip 2mm
\caption{Configuration of the fraction of particles surviving after
  $t=10^6$ unit times. About 2/3 of the initial number of particles have
  escaped after the time considered. The particles distributed on the
  bounds with larger indices have essentially retained their initial
  positions. The escape occured from bound $1$ only.} 
\label{fig.pos}
\end{figure}

Since the particles move with constant speed, in the limit where
$N\to\infty$ the particles will undergo
exponentially more frequent collisions as they move to the right-most end
of the lattice, and will be backscattered with exponentially increasing
probability, hence, in practice, never reaching the right boundary.
The escape is thus expected to be essentially due to exit from the left
boundary for large enough $N$. Figure \ref{fig.decay} shows the results
of a numerical simulation where the number of surviving particles is plotted 
vs. time. The power decay is
apparent at long times, with an asymptotic exponent whose value agrees
within a few percents 
with the value to be derived in Eq. (\ref{deltaval}).  
The configuration of the particles surviving after that time is 
displayed in Fig. \ref{fig.pos}. We note that the bonds are more or less
equally populated at the exception of the left-most bonds, which are
unpopulated.

%%%%%%%%%%%%%%%%%%%%%%%%%%%%%%%%%%%%%%%%%%%%%%%%%%%%%%%%%%%%%%%%%%%%%%%%%%%
\vskip 1cm
%%%%%%%%%%%%%%%%%%%%%%%%%%%%%%%%%%%%%%%%%%%%%%%%%%%%%%%%%%%%%%%%%%%%%%%%%%%

%%%%%%%%%%%%%%%%%%%%%%%%%%%%%%%%%%%%%%%%%%%%%%%%%%%%%%%%%%%%%%%%%%%%%%%%%%%
%%%%%%%%%%%%%%%%%%%%%%%%%%%%%%%%%%%%%%%%%%%%%%%%%%%%%%%%%%%%%%%%%%%%%%%%%%%

As discussed above Eq. (\ref{alg-decay}), the average (with
respect to random initial conditions) of the ratio of the number
of particles surviving after a time $t$ to their initial number $N_0$
defines the survival probability, which we would like to characterize in
terms of the two parameters of the persistent hierarchical graph, 
namely $\varepsilon$ and $\mu$.

%%%%%%%%%%%%%%%%%%%%%%%%%%%%%%%%%%%%%%%%%%%%%%%%%%%%%%%%%%%%%%%%%%%%%%%%
\section{The survival probability}
\label{sec.stay-prob}
%%%%%%%%%%%%%%%%%%%%%%%%%%%%%%%%%%%%%%%%%%%%%%%%%%%%%%%%%%%%%%%%%%%%%

The statistical average of a physical
observable $A(x_b)$
defined on the bonds of the graph is given by \cite{BG2001}
\begin{equation}
\langle A\rangle_t = \sum_{b=1}^{2B} \frac{1}{l_b} \int_0^{l_b} A(x_b) \;
\rho(x_b,t) dx_b =
\langle A \vert \hat P^t \rho_0 \rangle\ ,
\label{aver}
\end{equation}
where $\rho_0$ denotes the initial probability density which evolves with 
the Frobenius-Perron operator $\hat P^t$ to give, at time $t$, a density
$\rho(x_b,t)$ at position $x_b$ on the bond $b$.

In particular, if we consider the observable 
$A(x_b)=l_b$ for the bonds that compose the finite part of the graph 
and $A(x_c)=0$, for $c$ a scattering lead, Eq. (\ref{aver}) defines the
{\em survival probability}, i.~e.
%A convenient way of characterize the decay of an open system,
%is to consider 
the probability of finding the particle in the
interior of the system at a given time $t$,
\begin{equation}
P(t)=\sum_{b}\int_0^{l_b}\rho(x_b,t)dx_b\ .
\label{def-survival}
\end{equation}
We will henceforth reserve the notation $A$ for this observable.
One of our goals will be to show that this definition can indeed be decomposed 
as in Eq. (\ref{spectral2}).

Since we are interested in the time evolution at long times, we
may consider the spectral decomposition of $\hat P^t$ to get an asymptotic 
expansion valid for $t\to +\infty$ of the form 
\begin{equation}
P(t) = \langle A \vert \hat P^t \rho_0 \rangle = \sum_j
\langle A \vert
\Psi_j\rangle \ \mbox{e}^{s_jt} \ \langle
\tilde{\Psi}_j\vert \rho_0 \rangle + \dots
\label{spectral}
\end{equation}
as a sum of exponential functions\footnote{Possible extra terms such
as powers of the time multiplied by exponentials, $t^m\exp(s_jt)$,  
are not generic and may appear for particular values
of the parameters of the system.
%, that is lengths and scattering matrices at the vertices. 
See \cite{BG2001} for a discussion of this point.}.
Therefore Eq. (\ref{spectral2}) is obtained with 
\begin{equation}
A_j=\langle A |\Psi_j \rangle
\langle \tilde{\Psi}_j|\rho_0 \rangle\ .
\label{Aj}
\end{equation}

The spectral decomposition used in Eq. (\ref{spectral}) is fully determined
by the solutions of the problem \cite{BG2001}
\begin{equation}
{\mathsf Q}(s_j)\chi_j=\chi_j\ ,
\label{Qsj}
\end{equation}
which determine the Pollicott-Ruelle resonances $s_j$ and the corresponding
eigenstates $\chi_j$.
$\mathsf Q$ is a $2B \times 2B$ matrix with elements given by 
$Q_{bb'}(s)=P_{bb'}e^{-sl_{b'}}$.
%with $P_{bb'}$ defined in Eqs. (\ref{proba1}-\ref{proba2}).

Explicit expressions for the scalar products in Eq. (\ref{Aj}) were
found in \cite{BG2001}. For the right eigenstates,
\begin{equation}
\langle A\vert\Psi_j\rangle = \sum_b \chi_j[b] \; \frac{1}{l_b}
\int_0^{l_b} \mbox{e}^{-s_j
\frac{x_b}{v}} A(x_b) \; dx_b\ ,
\label{right.eigenstate}
\end{equation}
and for the left eigenstates,
\begin{equation}
\langle \tilde{\Psi}_j\vert\rho_0\rangle = \frac{1}{\sum_{b^{\prime\prime}}
l_{b^{\prime\prime}}
\tilde{\chi}_j[{b^{\prime\prime}}]^* \chi_j[{b^{\prime\prime}}]}
\; \sum_b \tilde{\chi}_j[{b^{\prime}}]^* \; \int_0^{l_{b^{\prime}}}
\mbox{e}^{s_j
\frac{x_{b^{\prime}}}{v}} \rho_0(x_{b^{\prime}}) \; dx_{b^{\prime}}\ .
\label{left.eigenstate}
\end{equation}
Here $\chi_j[b]$ denotes the $b$ component of the eigenstate $\chi_j$ and 
$\tilde{\chi}_j^*$ denotes the complex conjugate of the left eigenvector
of ${\mathsf Q}(s_j)$.
 
If for the initial density we take $\rho_0(x_b)=1$, for all bonds $b$, and 
$\rho_0(x_c)=0$ for infinite leads $c$,
that is a uniform distribution over the finite part of the graph, we have
\begin{equation}
A_j=\frac{1}{s_j^2}\frac
{\sum_{b,b'}\chi_j (b)\tilde{\chi}_j^*(b')
[e^{s_jl_{b'}}+e^{-s_jl_b}-e^{s_j(l_{b'}-l_b)}-1]}
{\sum_{b}l_b\chi_j (b)\tilde{\chi}_j^*(b)}
\ .
\label{Ajexp}
\end{equation}
As we will show in Sec. \ref{sec.pert.th}, the Pollicott-Ruelle resonances
$s_j$ are small so that we can expand the exponential terms in
Eq. (\ref{Ajexp}) and get, to first order,
\begin{equation}
A_j=\frac{\sum_b l_b \chi_j[b] \sum_{b}l_b \tilde{\chi}_j^*[b]}
{\sum_b l_b \chi_j[b]\tilde{\chi}_j^*[b]}\ .
\label{Aj2}
\end{equation}

%%%%%%%%%%%%%%%%%%%%%%%%%%%%%%%%%%%%%%%%%%%%%%%%%%%%%%%%%%%%%%%%%%%%%%%
\section{Pollicott-Ruelle resonances}
\label{sec.pert.th}
%%%%%%%%%%%%%%%%%%%%%%%%%%%%%%%%%%%%%%%%%%%%%%%%%%%%%%%%%%%%%%%%%%%%%%%
According to Eq. (\ref{Qsj}), the Pollicott-Ruelle resonances $s_j$ are the
roots of the following determinant,
\begin{equation}
\det\big[{\mathsf I} - {\mathsf Q}(s)\big]=0\ .
\label{detImQ}
\end{equation}
In order to write explicitly the matrix ${\mathsf Q}$, we will order the
states according to
\begin{equation}
\left(1,+\quad 1,-\quad 2,+\quad 2,-\quad \cdots\quad N,+\quad N,-
\right)\ .
\label{order}
\end{equation}

This way ${\mathsf Q}$ has the expression
\begin{equation}
{\mathsf Q}(s) = 
\left(
\begin{array}{c@{\quad}c@{\quad}c@{\quad}c@{\quad}c@{\quad}c@{\quad}c@{\quad}c}
0&q_1 e^{-s l_1}&0&0&0&\dots&0&0\\
q_2 e^{-s l_1}&0&0&p_2 e^{-s l_2}&0&\dots&0&0\\
p_2 e^{-s l_1}&0&0&q_2e^{-s l_2}&0&\dots&0&0\\
0&0&q_3e^{-s l_2}&0&0&\dots&0&0\\
0&0&p_3e^{-s l_2}&0&0&\dots&0&0\\
\vdots&\vdots&\vdots&\vdots&\vdots&\ddots&\vdots&\vdots\\
0&0&0&0&0&\dots&q_{N+1}e^{-s l_{N}}&0
\end{array}
\right)\ .
\label{expQ}
\end{equation}
Since this is a sparse matrix, it is rather straightforward
to compute the determinant Eq. (\ref{detImQ}),
\begin{equation}
\det\big[{\mathsf I} - {\mathsf Q}(s)\big]=\prod_{i=1}^N\delta_i\ ,
\label{detprod}
\end{equation}
where 
\begin{eqnarray}
\delta_i &=& 1 - q_i q_{i+1}e^{-2 s l_i}\left[1+\left(\frac{p_i}{q_i}\right)^2
\left(\frac{1}{\delta_{i-1}}-1\right)\right]\ , \quad i\ge 2\
,\label{deltai}\\
\delta_1&=&1 - q_1q_2e^{-2 s l_1}\ .\label{delta1}
\end{eqnarray}
Owing to the product structure of Eq. (\ref{detprod}), the zeros of
Eq. (\ref{detImQ}) are the zeros of $\delta_N$. One can compute them 
numerically for any value of the parameters $\varepsilon$ and
$\mu$. It should be emphasized that the identification of zeros in 
Eqs. (\ref{detprod}-\ref{delta1}) holds only for finite $N$.
However, the numerical resolution of Eqs. (\ref{deltai}-\ref{delta1}) is
limited to small $N$ and,
for the sake of proving Eq. (\ref{delta}), a perturbative approach
allows an analytic treatment. This is done in what follows.

%%%%%%%%%%%%%%%%%%%%%%%%%%%%%%%%%%%%%%%%%%%%%%%%%%%%%%%%%%%%%%%%%%%%
\subsection{Perturbation Theory}
\label{subsec.pert.th}
%%%%%%%%%%%%%%%%%%%%%%%%%%%%%%%%%%%%%%%%%%%%%%%%%%%%%%%%%%%%%%%%%%%%

In order to set up a perturbation scheme, we choose $\varepsilon$ as our
small parameter and
note that the ``unperturbed'' system,
$\varepsilon=0$, corresponds to the union of $N$ non-interacting bonds.
That is, the particles are oscillating back and forth on the same bond.
The spectrum of this unperturbed system is the union of the 
$s_{n,m}=i\frac{m\pi}{l_n}$, $m \in \mathbb{Z}$, 
$n=1,\ldots,N$.
The resonances with $s \neq 0$ are not degenerate and remain isolated under
the perturbation, even in the limit $N\to\infty$. After the perturbation,
each isolated resonance adds one contribution to Eq. (\ref{spectral}), with 
an $s_j$ whose real part is negative and $\mathcal{O}(\epsilon)$. Therefore
the states associated to them decay exponentially fast
(with an oscillation on top).  
On the other hand, the resonance $s=0$ is the only degenerate unperturbed
resonance, with a  
multiplicity $N$. As it will turn out, the perturbation acting on this
resonance reduces the degeneracy by one unit at each order of the
perturbation, the splitting being proportional to $\epsilon^j$ with $j$ 
the order of the perturbation. Thus, in the limit $N\to\infty$, the
spectrum has an accumulation point at $s=0$. Therefore
these states cannot be considered isolated and their contribution to 
Eq. (\ref{spectral}) must be accounted for separately
from that of the isolated resonances because it becomes an integral
in this limit. 
This integral accounts for the algebraic decay of the surviving
       probability in the long time limit.

Let us discard the isolated resonances and consider only the resonances 
$s_{n,0}$. The eigenstates associated to the unperturbed system
are solutions of the equation
\begin{equation}
{\mathsf Q}^{(0)}(0)\chi^{(0)}_n = \chi^{(0)}_n\ ,
\label{Q0chi0}
\end{equation}
where ${\mathsf Q}^{(0)}$ is given by Eq. (\ref{expQ}), in which $\varepsilon$ 
is set to zero. Explicit expressions for the eigenvectors are~:
\begin{equation}
\chi^{(0)}_1 = \frac{1}{\sqrt{2}}
\left(\begin{array}{c}1\\1\\0\\0\\0\\\vdots\\0\end{array}\right)\ ,\quad
\chi^{(0)}_2= \frac{1}{\sqrt{2}}
\left(\begin{array}{c}0\\0\\1\\1\\0\\\vdots\\0\end{array}\right)\ ,
\dots ,\quad
\chi^{(0)}_N= \frac{1}{\sqrt{2}}
\left(\begin{array}{c}0\\0\\\vdots\\0\\ç0\\1\\1\end{array}\right)
\ .
\label{psi0}
\end{equation}

In order to implement the perturbation theory in powers of $\varepsilon$, we
first consider the right eigenvectors. The calculation transposes
straightforwardly to the case of left eigenvectors. 
The perturbation theory closely resembles the standard perturbation theory
for degenerate eigenvalues \cite{Landau3}.
Let us consider linear combinations
\begin{equation}
\chi = \sum_{i=1}^N c_{i} \chi^{(0)}_i\ ,
\label{psin}
\end{equation}
where the coefficients $c_{i}$ are polynomials in $\varepsilon$, and seek 
approximate solutions of the system
\begin{equation}
{\mathsf Q}(s) \chi = \chi\ ,
\label{Qnpsin}
\end{equation}
where $s$ is a polynomial in $\varepsilon$ and
${\mathsf Q}$ will be expanded to a given order in $\varepsilon$. Writing
${\mathsf Q} = {\mathsf Q}^{(0)} + \delta{\mathsf Q}$, we substitute
Eq. (\ref{psin}) into Eq. (\ref{Qnpsin}). 
Multiplying both sides of Eq. (\ref{Qnpsin}) by
$\chi_i^{(0)}$, $i=1,\dots,N$, and using Eq. (\ref{Q0chi0}), we obtain a  
system of $N$ linear equations for the coefficients $c_{i}$,
$\sum_{j=1}^N V_{i,j}(s)c_{j} = 0$,
where $V_{i,j}(s)={\chi_i^{(0)}}^{\rm T}
[{\mathsf Q}(s)-{\mathsf Q}^{(0)}(0)]\chi_j^{(0)}$
are the matrix elements of the perturbation operator 
\begin{equation}
{\mathsf V}_1(s)= \frac{1}{2}
\left(
\begin{array}{c@{\quad}c@{\quad}c@{\quad}c@{\quad}c}
\scriptstyle
-2 + e^{-s l_1}(q_1+q_2)&\scriptstyle e^{-s l_2}p_2&
0&\dots&0\\
\scriptstyle e^{-s l_1}p_2&\scriptstyle -2 + 
e^{-s l_2}(q_2+q_3)&
\scriptstyle e^{-s l_3}p_3&\dots&0\\
0&\scriptstyle e^{-s l_2}p_3&\scriptstyle -2 + 
e^{-s l_3}(q_3+q_4)&\dots&0\\
\vdots&\vdots&\vdots&\ddots&\vdots\\
0& 0&0&\dots&
\scriptstyle -2 + e^{-s l_N}(q_N+q_{N+1})
\end{array}
\right)
\label{V}
\end{equation}
The values of $s$ are found by solving the secular equation
\begin{equation}
\det[{\mathsf V}_1(s)] =0
\label{seceqV}
\end{equation}
to the desired power in $\varepsilon$.

Expanding $p_n$ and $q_n$ in powers of $\varepsilon$, we can compute
the corrections to the unperturbed solution. 
In fact, expanding Eq. (\ref{seceqV}) up to ${\cal O}(\varepsilon)$ 
we find that only one eigenstate, $s_{1,0}$, has negative real part, 
\begin{equation}
s_{1,0}= -\frac{p_0 \varepsilon}{2 l_1} +
{\cal O}(\varepsilon^2)\label{1stordersigma1}\ ,
\end{equation}
while up to this order, the others remain degenerate, 
\begin{equation}
s_{n,0} = {\cal O}(\varepsilon^2)\ , \quad n\ge2\ .\label{1stordersigmai}
\end{equation}
The eigenvector corresponding to $s_{1,0}$ is
\begin{equation}
\chi(s_1) = \chi_1^{(0)} + {\cal O}(\varepsilon)\ .
\label{1storderpsi1}
\end{equation}
Hence the degeneracy remains to be lifted among the $N-1$ remaining
eigenmodes. 
We study now how the second order correction affects the degenerate state.
We proceed in a similar manner as we did for the first order. The only  
difference
is that now $\chi_1^{(0)}$ does not belong to the base of the degenerate 
subspace.
Accordingly the perturbation operator in this subspace is 
represented
by the matrix ${\mathsf V}_2=(V_{i,j})_{2\le i,j\le N}$, which is obtained
from the matrix ${\mathsf V}_1$ by
removing the first line and first column. Expanding the equation 
$\det[{\mathsf V}_2]=0$
up to ${\cal O}(\varepsilon^2)$, we get a result similar to
Eqs. (\ref{1stordersigma1}-\ref{1storderpsi1}) with $s_{2,0} = -p_0
\varepsilon^2/2 l_2+{\cal O}(\varepsilon^3)$ and 
$\chi_2 = \chi^{(0)}_2 +{\cal O}(\varepsilon^3).$
Proceeding, the effect of the perturbation at the third order in $\varepsilon$
must be studied among the remaining $N-2$ degenerate states. By induction we  
thus have a whole hierarchy of roots, each
corresponding to a different order in $\varepsilon$ and determined by
secular equations involving the corresponding perturbation operator that
acts in the still degenerate subspace
${\mathsf V}_n = (V_{i,j})_{n\le i,j\le N}$ expanded up to 
${\cal O}(\varepsilon^n)$. 
It is clear that, at any given order of the perturbation theory, the 
resonances which are not anymore part of the degenerate subspace will have
further corrections to their values. However we do not need to take them
into consideration since we are only interested in the leading
contributions to every resonance of the spectrum.
In fact we can prove the 

\begin{proposition}
\label{prop.si}
The $N$ roots of Eq. (\ref{detImQ}) can be approximated to order $N$ in
$\varepsilon$ by $s_{1,0},\dots,\,s_{N,0}$, where, for every $1\le n\le
N$, $s_{n,0}$ is the only root of order $\varepsilon^n$ of the secular
equation
\begin{equation}
\det[{\mathsf V}_n(s)]=0\ ,
\label{detVi}
\end{equation}
with leading contribution 
\begin{equation}
s_{n,0}=-\frac{p_0 \varepsilon^n}{2 l_n}+{\cal O}(\varepsilon^{n+1}).
\label{si0}
\end{equation}
The corresponding right-eigenvector $\chi$, Eq. (\ref{psin}), has coefficients 
$c_n,\ldots,\,c_N$ which are the solutions of the linear system
\begin{equation}
\sum_{k=n}^NV_{j,k}(s_{n,0})c_k = 0\ ,\quad j\ge n\ ,
\label{cksigmai}
\end{equation}
Left-eigenvectors $\tilde{\chi}$ have coefficients determined by
\begin{equation}
\sum_{j=n}^NV_{j,k}(s_{n,0})c_j = 0\ ,\quad k\ge n\ .
\label{leftcksigmai}
\end{equation}
\end{proposition}
We will not discuss the states associated to $s_{n,m}$ ($m\neq 0$),
which are exponentially decaying states and play a role only at the early
stages of the dynamics.

From Eqs. (\ref{cksigmai}-\ref{leftcksigmai}) it is
easy to show that to the leading order we have 
\begin{equation}
\chi_n=\tilde{\chi}_n=\chi^{(0)}_n
+{\cal O}(\varepsilon^n)\ .
\label{chinOn}
\end{equation}

%%%%%%%%%%%%%%%%%%%%%%%%%%%%%%%%%%%%%%%%%%%%%%%%%%%%%%%%%%%%%%%%%%%%%%%%%%%
\subsection{Algebraic Decay}
\label{sub.sec.alg.dec}
%%%%%%%%%%%%%%%%%%%%%%%%%%%%%%%%%%%%%%%%%%%%%%%%%%%%%%%%%%%%%%%%%%%%%%%%%%%
According to Prop. \ref{prop.si}, the resonances $s_{n,0}$ are 
${\mathcal O}(\varepsilon^n)$ and therefore accumulate to $s=0$ as $n$ becomes
large, thus proving what we promised. Equation (\ref{Aj2}) 
together with Eq. (\ref{chinOn}) and the expression of the unperturbed
eigenvectors, Eq. (\ref{psi0}), allow us to evaluate the leading 
contribution to $A_j$~:
$A_j = 2 \mu^j +{\cal O}(\varepsilon)$,
where we have substituted $l_j=l_0 \mu^j$ and $l_0=1$.

Turning back to Eq. (\ref{delta}), we have shown that the decay is
algebraic as in Eq. (\ref{alg-decay}) with 
\begin{equation}
\delta = \frac{1}{\ln \varepsilon/\ln \mu -1}\ .
\label{deltaval}
\end{equation}

We point out as a conclusion to this section that both parameters of the
persistent hierarchical graph,
$\epsilon$ and $\mu$, are necessary to grant the algebraic decay of the
survival probability. This point will be further discussed in the conclusions.
In what follows we will derive further properties of
the persistent hierarchical graphs, first classical and then quantum.

%%%%%%%%%%%%%%%%%%%%%%%%%%%%%%%%%%%%%%%%%%%%%%%%%%%%%%%%%%%%%%%%%%%%%%%%%%%%
\section{Thermodynamic formalism}
\label{sec.pres}
%%%%%%%%%%%%%%%%%%%%%%%%%%%%%%%%%%%%%%%%%%%%%%%%%%%%%%%%%%%%%%%%%%%%%%%%%%%%

For the real time process we consider, the free energy (usually referred to 
as topological pressure) per unit time is defined in analogy to continuous
time processes where the stretching factors are here replaced by the
inverses of the transition probabilities at the vertices of the graph
\cite{BG2001}~: 
\begin{equation}
\C.F(\beta) = \lim_{T\to\infty}\frac{1}{T}\ln \C.Z_T(\beta)\ ,
\label{fedef}
\end{equation}
where the dynamical partition function
\begin{equation}
\C.Z_T(\beta) = \sum_{\underline{b}_T} 
[P_{b_0b_1}\cdots P_{b_{n-1}b_{n}}]^\beta
\label{pfdef}
\end{equation}
is the sum over all trajectories of time length $T$ (or equivalently length
$L=vT$) of their respective
probabilities raised to the power $\beta$, with $\beta>0$ playing the
role of an inverse temperature. We note that the length of the trajectories 
cannot be measured sharply because of the continuous time nature of the
system. Rather the sum in Eq. (\ref{pfdef}) should be understood as a sum
over all trajectories whose lengths are within an interval $T\pm\Delta t$,
where $\Delta t$ is fixed. In the infinite $T$ limit, the value of $\Delta
t$ is irrelevant.

Some properties of the free energy, are \cite{GD95}~: (i) 
$\C.F$ is a monotonically decreasing function of $\beta$; (ii) $\C.F$ has a 
zero for some $\beta=d_H$, $0<d_H<1$ (the strict 
inequality being due to the open boundaries), where (iii)
$d_H$ is the fractal dimension of the repeller with respect to a
properly defined metric space (in the sense that the space of trajectories
is a continuum where trapped trajectories form a subset with fractal
dimension); (iv) for hyperbolic systems of one degree of
freedom, $-\C.F'(d_H)$ is the value of the positive Lyapunov exponent
on the space of trapped trajectories; (v) $-\C.F(1)$ measures the rate of escape
from the system; (vi) the difference between these last two quantities is
the metric (Kolmogorov-Sinai) entropy on the repeller, 
$h_{\mathrm KS}=\C.F(1)-\C.F'(d_H)$; 
and (vii) $\C.F(0)\equiv h_{\mathrm{TOP}}$ is the topological entropy. 

As argued in \cite{BG2001}, the free energy Eq. (\ref{fedef}) can be
obtained as the leading zero of the following zeta function
$\zeta(s,\beta) = \det[\mathsf{I} - \mathsf{Q}_\beta(s)]$,
where $\mathsf{Q}_\beta$ is identical to the matrix $\mathsf{Q}$ defined 
in Eq. (\ref{expQ}), with the probabilities $q_i$ and $p_i$ now raised to
the power $\beta$. Thus the free energy $\C.F(\beta)$ is the leading solution
$s$ of the expression
$\prod_i \delta_i = 0$,
where $i$ takes values on the set of bonds and 
the $\delta_i$ are determined by the recurrence relation
\begin{eqnarray}
\delta_i &=& 1 - q_i^\beta q_{i+1}^\beta e^{-2 s l_i}\left[1 + \left(\frac
{p_i}{q_i}\right)^{2\beta}\left(\frac{1}{\delta_{i-1}}-1\right)\right]\ ,
\label{deltaibeta}\\
\delta_1 &=& 1 - q_1^\beta q_{2}^\beta e^{-2 s l_1}\ .
\label{delta1beta}
\end{eqnarray}

We prove the following asymptotic behaviors~:
\begin{proposition}
\label{largebeta}
In the limit of large $\beta$, the free energy $\C.F$ is linear in $\beta$
with a coefficient exponentially small with respect to the number of
bonds in the system, $N$, 
\begin{equation}
\lim_{\beta\to\infty}\C.F(\beta) = -\beta\frac{1+\varepsilon}{2}
\left(\frac{\varepsilon}{\mu}\right)^N\ .
\label{Flimbetainf}
\end{equation}
\end{proposition}
The proof of this result follows by considering the probability of a particle 
bouncing off a given bond $n$ for a time $T$. Let $W_T(n)$ be this
probability. We have
\begin{eqnarray}
W_T(n) &=& [q_n q_{n+1}]^{T/2\mu^n}\ ,\nonumber\\
&=& \left[1 - \frac{1 +\varepsilon}{2}\left(\frac{\varepsilon}{\mu}\right)^n 
+ O\left(\frac{\varepsilon^2}{\mu}\right)^n\right]^T\ .
\label{WTn}
\end{eqnarray}
Thus the ratio 
\begin{equation}
\frac{W_T(n)}{W_T(n-1)} = \left[1 + \frac{1+\varepsilon}{2}\left(1 -
    \frac{\varepsilon}{\mu}\right)
  \left(\frac{\varepsilon}{\mu}\right)^{n-1}
+\ldots\right]^T \quad>1
\label{ratioWTn}
\end{equation}
is larger than 1, which implies that, as $\beta\to\infty$, the free energy
is dominated by particles bouncing off the last
bond, i.~e. $F(\beta)\approx \lim_{T} (1/T) \ln [W_T(N)^\beta]$. 
Equation (\ref{Flimbetainf}) follows.

\begin{figure}
\epsfxsize=12truecm
\epsfysize=8truecm
\epsfbox{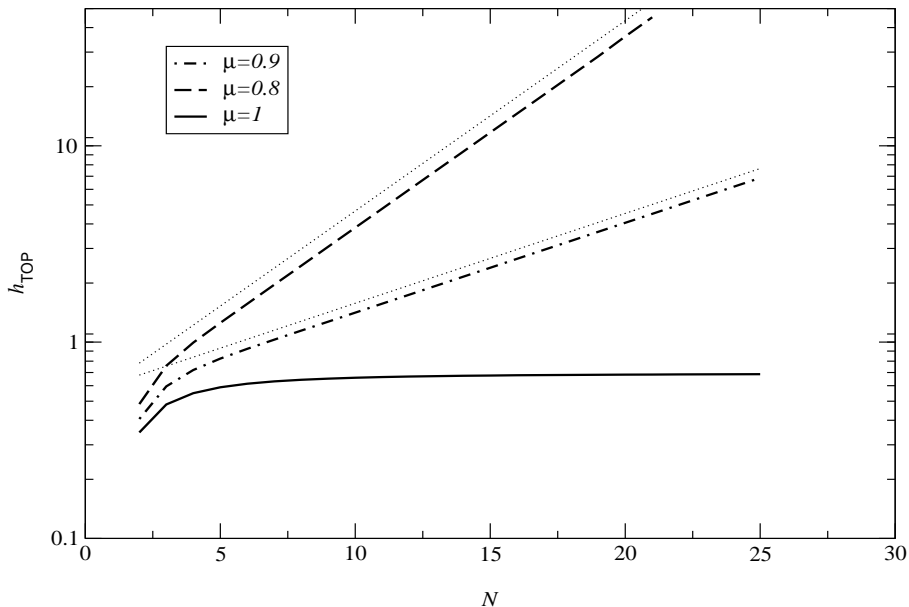}
\vskip 2mm
\caption{Topological entropy vs. $N$ for three different values of $\mu$~:
  $\mu=1$ (solid line), $9/10$ (dot-dashed line) and $8/10$ (dashed
  line). The two dotted curves  are proportional to $1/\mu^N$.}
\label{fightop}
\end{figure}

\begin{proposition}
\label{smallbeta}
When $\beta$ tends to zero, the free energy $\C.F$ has a limit independent of 
$\varepsilon$, given by
$\lim_{\beta\to 0}\C.F(\beta)\propto\frac{1}{\mu^N}[1 + O(\mu)]\ .$
Thus in the limit of large number of bonds $N$, the free energy has a 
singular limit,
$\lim_{N\to\infty}\C.F(0)=\infty\ .$
\end{proposition}
This holds since the rate of creation of new trajectories per collision is
the same at every site, while the rate of collision per unit time increases
exponentially as the particles go deeper into the lattice. Hence the
topological entropy is overwhelmingly dominated by particles bouncing off
the right-most bond. In particular, the reason for its diverging with 
$N\to\infty$ is due to the existence of trajectories undergoing an infinite 
number of collisions in a finite time, e.~g. the trajectory ever moving to
the right. Numerical evidence for Prop. \ref{smallbeta} is shown in Fig.
\ref{fightop}.
%%%%%%%%%%%%%%%%%%%%%%%%%%%%%%%%%%%%%%%%%%%%%%%%%%%%%%%%%%%%%%%%%%%%%%%%%%%
%%%%%%%%%%%%%%%%%%%%%%%%%%%%%%%%%%%%%%%%%%%%%%%%%%%%%%%%%%%%%%%%%%%%%%%%%%%

We close this section with the observation that, in the infinite
system limit, we expect the free energy to have a phase transition at the 
value $\beta=1$. This should result from the
asymptotic behaviors of the free energy discussed in Props.
\ref{largebeta} and \ref{smallbeta}.
In the limit $N\to\infty$, by Prop. \ref{largebeta}, 
the free energy is zero for every $\beta>1$ since it is zero at
$\beta\to\infty$. On the other hand, by
Prop. \ref{smallbeta}, it diverges as $\beta\to 0$, and thus must decrease
steeply for $0<\beta<1$. We thus infer that matching the two curves at 
$\beta=1$ results in a discontinuity of one of the derivatives of $\C.F$.

%%%%%%%%%%%%%%%%%%%%%%%%%%%%%%%%%%%%%%%%%%%%%%%%%%%%%%%%%%%%%%%%%%%%%%%%%%%%%
\section{The quantum hierarchical graph}
\label{sec.qres}
%%%%%%%%%%%%%%%%%%%%%%%%%%%%%%%%%%%%%%%%%%%%%%%%%%%%%%%%%%%%%%%%%%%%%%%%%%%%%

In this section we wish to explore the possibility of having a scaling
relation for the widths of the resonances, as was suggested in
\cite{ketzmerick}. Numerical analysis of the time evolution of quantum
systems as considered in \cite{BGsub,jordan} is possible for a finite
system, but goes beyond our scope. Thus we consider a quantum system whose
classical limit is 
the one defined in Sec. \ref{sec.model}. The quantum system is a linear
chain with transition and reflection {\em probability amplitudes} 
\begin{equation}
\left\{
\begin{array}{lcl}
\sigma_{(n+1,+),(n,+)}&=&\sqrt{p_{n+1}}\ , \\
\sigma_{(n,-),(n,+)}&=&i \sqrt{q_{n+1}}\ , \\
\sigma_{(n-1,-),(n,-)}&=&\sqrt{p_{n}}\ , \\
\sigma_{(n,+),(n,-)}&=&i \sqrt{q_{n}}\ ,
\end{array}
\right.
\label{qampl}
\end{equation}
with $p_n$ and $q_n$ defined as in Eq. (\ref{pn}).
$\sigma_{bb'}=0$ for all other possibilities.
It is clear that at each vertex the scattering matrix 
is unitary and therefore the quantum 
problem is well defined. Moreover we have that 
$P_{bb'}=|\sigma_{bb'}|^2$ which shows \cite{BGsub} that the classical 
limit of this quantum problem is indeed given by
Eqs. (\ref{proba1}-\ref{proba2}).

The time evolution of a wave packet in an open system is controlled
by the scattering resonances defined in the complex plane of wavenumbers
$k$ as the poles of the scattering matrix. For a quantum system, denoting
the wavefunction by $\psi(x_b,t)$, we can write the survival probability as~:
$P_{\scriptscriptstyle \mathrm QM}(t) = \sum_{b}\int dx_b |\psi(x_b,t)|^2$.
Letting
$\psi(x_b,t) = \sum_{r}c_r(x_b)e^{-i E_r t}$,
where $c_r(x_b)$ are determined by eigenstates of the evolution operator
with complex eigenvalues $E_r$, we can rewrite the survival probability
as follows~:
\begin{equation}
P_{\scriptscriptstyle\mathrm QM}(t) = \sum_{r, r'}c_r c_{r'} e^{i(\epsilon_r -
  \epsilon_{r'})t} e^{-(\Gamma_r+\Gamma_{r'})t/2}\ ,
\label{qusurproba2}
\end{equation}
where we have used the decomposition $E_r = k_r^2=\epsilon_r - i\Gamma_r/2$.

Since for short times the quantum evolution follows the classical one,
it is interesting to study the distribution of scattering resonances in
hierarchical graphs and look for manifestations of the algebraic decay
in a quantum spectrum. Hufnagel {\em et al.} \cite{ketzmerick}
showed, in the framework of a quantum version of the chain model, that the 
distribution $p$ of the width $\Gamma=-4{\mathrm Re}\,k{\mathrm Im}\, k$ of 
the quantum scattering resonances $k$ 
satisfies $p(\Gamma) \sim 1/\Gamma$, $\Gamma\ll 1$. Moreover, using an 
argument based on perturbation theory, they argued
that the imaginary parts of the quantum scattering resonances
satisfy a scaling relation.

Given a scaling relation for the resonance widths, $\Gamma_i=f^i$,
the widths distribution follows:
\begin{equation}
p(\Gamma)=\sum_{i}\delta(\Gamma_i-\Gamma) \approx
\int\delta(f^x-\Gamma)dx \sim
\frac{1}{\Gamma}
\label{feq}
\end{equation}

This distribution has been associated to peaks that decorate
fractal conductance fluctuations observed in energy scales 
larger than the mean level spacing.
But since the width and height of the peaks in the conductance are determined
by the imaginary part of the scattering resonances,
the scaling behavior of the resonance widths is also contributing to the
self similar, i.~e. fractal, shape of the conductance.

The scattering resonances are the zeros of the zeta function
$\det[\mathsf{I}-\mathsf{R}(k)]\ ,$
with the matrix $\mathsf{R}(k)$
obtained by replacing $s$ by $i k$ and $P_{bb'}$ by $\sigma_{bb'}$
in Eq. (\ref{expQ}). As for the classical resonances,
we can develop a perturbative approach in order to determine the quantum 
resonances and the corresponding eigenstates, $\mathsf{R}(k) \phi=\phi$.

As opposed to the classical case, the zeroth order resonances 
are generally non-degenerate. Indeed a straightforward calculation of
$\det[\mathsf{I}-\mathsf{R}(k)]$ with $\varepsilon=0$ yields the roots
\begin{equation}
k^{(0)}_{n,p}=\frac{(2 p + 1)\pi}{2 l_n}\ ,\quad p\in{\mathbb Z}\ . 
\label{qures0}
\end{equation}
Similarly to Eq. (\ref{psi0}), the eigenvector corresponding to 
$k_{n,p}$ is given by
\begin{equation}
\phi_{n,p}=
(0\quad \dots\quad 0\quad \overbrace{1\quad (-1)^p}^{\hspace{-5mm}
2n - 1\quad2n}
\quad 0\quad \dots\quad 0)\ .
\label{phinp}
\end{equation}

Unless $\mu=1$, the zeroth order quantum resonances are all isolated, 
whereas, for
$\mu=1$, $k^{(0)}_{n,p}=(2p+1)\pi/2$ is independent of $n$, so that, for every 
different $p$, the resonances have an $N$th order degeneracy. 
We consider the two different cases separately.

%%%%%%%%%%%%%%%%%%%%%%%%%%%%%%%%%%%%%%%%%%%%%%%%%%%%%%%%%%%%%%%%%%%%%%%%%%%
\subsection{Non-hierarchical Graph~: $\mu=1$}
%%%%%%%%%%%%%%%%%%%%%%%%%%%%%%%%%%%%%%%%%%%%%%%%%%%%%%%%%%%%%%%%%%%%%%%%%%%

For $\mu=1$, the situation is similar to Sec. \ref{sec.pert.th}. We can
proceed by analogy and show the following

\begin{proposition}
For equally inter spaced scatterers, i.~e. $\mu=1$, the quantum resonances
are given by 
\begin{equation}
k_{n,p}= \frac{(2 p +1)\pi}{2} - i\frac{p_0 \varepsilon^n}{4} + {\mathcal O}
(\varepsilon^{n+1})\ ,\quad p\in{\mathbb Z}\ .
\quad\quad (\mu=1)
\label{mu1knp}
\end{equation}
\end{proposition}
Thus, for every integer $p$, the $k_{n,p}$ satisfy a scaling law and we
have an accumulation point at $(2p+1)\pi/2$.
Since the inverse of the lifetime is 
$\Gamma =-4 \mathrm{Re} \,k \, \mathrm{Im} \,k$
and $v=2\mathrm{Re} \,k$ is identified with the speed of the particle 
we have that $\Gamma_{n,p}=v\frac{p_0}{2}\varepsilon^j=s_j$.

This result shows that asymptotically the classical and quantum lifetimes
of the resonances with longest lifetime coincide in the graph with evenly
inter spaced scatterers ($\mu=1$). This is in opposition
to fully chaotic graphs where the strict inequality is satisfied 
$\gamma > \Gamma_{min}$, \cite{GaspRice}. However there is no algebraic
decay in this case.

%%%%%%%%%%%%%%%%%%%%%%%%%%%%%%%%%%%%%%%%%%%%%%%%%%%%%%%%%%%%%%%%%%%%%%%%%%%
\subsection{Hierarchical Graph~: $\mu\neq 1$}
%%%%%%%%%%%%%%%%%%%%%%%%%%%%%%%%%%%%%%%%%%%%%%%%%%%%%%%%%%%%%%%%%%%%%%%%%%%

The case $\mu\neq 1$ is trickier. Indeed, one expects that the lowest order 
correction to $k_{n,p}$ is ${\mathcal O}(\varepsilon^n)$, but it can only be 
determined in the perturbation theory provided we know the $n-1$th order
correction to the roots $k_{n',p'}$ with $n'<n$, as well as the corresponding 
eigenvectors. In the remaining of this section, we will outline the derivation
of  the first order resonances and their eigenvectors and present in table
\ref{tab.knp} the results of a computation of $k_{n,p}$ to fourth order of 
perturbation theory.

\begin{table}
\begin{tabular}{||c|c|*{4}{|c}||}
\hline
\hline
$n$&$p$
&
${\mathcal O}(\varepsilon^0)$
&
${\mathcal O}(\varepsilon^1)$
&
${\mathcal O}(\varepsilon^2)$
&
${\mathcal O}(\varepsilon^3)$
\\
\hline
$1$&$1$&
$\frac{\pi}{2 \mu} $
&
$-\frac{i p_0}{4 \mu}$
&
$-\frac{i p_0 [2 -2 e^{i \pi \mu} + p_0 + e^{i \pi \mu}p_0]}
{8 [1 + e^{i \pi \mu}]\mu} $
&
$-i\frac{[1 + e^{i \pi \mu}]^2 p_0^3 - 
3 e^{i \pi \mu} p_0^2 \mu}{12 [1 + e^{i \pi \mu}]^2 \mu}$
\\
$1$&$0$&
$-\frac{\pi}{2 \mu} $
&
$-\frac{i p_0}{4 \mu}$
&
$-\frac{i p_0 [-2 +2 e^{i \pi \mu} + p_0 + e^{i \pi \mu}p_0]}
{8 [1 + e^{i \pi \mu}]\mu} $
&
$-i\frac{[1 + e^{i \pi \mu}]^2 p_0^3 - 
3 e^{i \pi \mu} p_0^2 \mu}{12 [1 + e^{i \pi \mu}]^2 \mu}$
\\
$2$&$1$&
$\frac{\pi}{2 \mu^2}$
&
$0$
&
$i\frac{[-1 + e^{i \pi/\mu}]p_0}{4 [1 + e^{
i \pi\/\mu}] \mu^2}$
&
$-i\frac{-[1 + e^{i \pi/\mu}]^2 [-1 + e^{
i \pi \mu}] p_0 + e^{i \pi/\mu} [1 + e^{i \pi \mu}]p_0^2}
{4 [1 + e^{i \pi/\mu}]^2 [1 + e^{i \pi \mu}] \mu^2}
$
\\
$2$&$0$
&
$-\frac{\pi}{2 \mu^2}$
&
$0$
&
$-i\frac{[-1 + e^{i \pi/\mu}]p_0}{4 [1 + e^{
i \pi\/\mu}] \mu^2}$
&
$-i\frac{[1 + e^{i \pi/\mu}]^2 [-1 + e^{
i \pi \mu}] p_0 + e^{i \pi/\mu} [1 + e^{i \pi \mu}]p_0^2}
{4 [1 + e^{i \pi/\mu}]^2 [1 + e^{i \pi \mu}] \mu^2}
$
\\
$3$&$1$&
$\frac{\pi}{2 \mu^3}$
&
$0$
&
$0$
&
$i\frac{[-1 + e^{i \pi/\mu}] p_0}
{4 [1 + e^{i \pi/\mu}] \mu^3}$
\\
$3$&$0$&
$-\frac{\pi}{2 \mu^3}$
&
$0$
&
$0$
&
$-i\frac{[-1 + e^{i \pi/\mu}] p_0}
{4 [1 + e^{i \pi/\mu}] \mu^3}$
\\
$4$&$1$&
$\frac{\pi}{2 \mu^4}$
&
$0$
&
$0$
&
$0$
\\
$4$&$0$&
$-\frac{\pi}{2 \mu^4}$
&
$0$
&
$0$
&
$0$
\\
\hline
\end{tabular}
\vskip 1cm
\begin{tabular}{||c|c||c||}
\hline
$n$&$p$&
${\mathcal O}(\varepsilon^4)$\\
\hline
$1$&$1$&
$-i\frac{
[1 + e^{i \pi \mu}]^3 p_0^4 - 
2 [-1 + e^{i \pi \mu}] p_0^2 [1 + e^{2 i \pi \mu} - 
2 e^{i \pi \mu} (-1 + \mu)] + e^{i \pi \mu} p_0^3 [-2 + e^{i \pi \mu} 
(-2 + \mu) - \mu] \mu}{16 [1 + e^{i \pi \mu}]^3 \mu}$
\\
$1$&$0$&
$-i\frac{[1 + e^{i \pi \mu}]^3 p_0^4 + 
2 [-1 + e^{i \pi \mu}] p_0^2 [1 + e^{2 i \pi \mu} - 
2 e^{i \pi \mu} (-1 + \mu)] - e^{i \pi \mu} p_0^3 [2 + e^{i \pi \mu} 
(2 + \mu) - \mu] \mu}
{16 [1 + e^{i \pi \mu}]^3 \mu}$
\\
$2$&$1$&
$i \frac{p_0^2 [- 2 \mu+ (4-2\mu-p_0\mu) e^{i \pi/\mu} - (4-p_0\mu+3p_0\mu)
e^{2 i \pi/\mu}  + 2\mu e^{3 i \pi\mu}]}{16
[1 + e^{i \pi/\mu}]^3 \mu^3}$
\\
$2$&$0$&
$-i \frac{p_0^2 [- 2 \mu+ (4-2\mu+3p_0\mu) e^{i \pi/\mu} - (4-p_0\mu-p_0\mu)
e^{2 i \pi/\mu}  + 2\mu e^{3 i \pi\mu}]}{16
[1 + e^{i \pi/\mu}]^3 \mu^3}$
\\
$3$&$1$&
$i\frac{[-1 + e^{i \pi \mu}] p_0}
{4 [1 + e^{i \pi \mu}] \mu^3)}$
\\
$3$&$0$&
$-i\frac{[-1 + e^{i \pi \mu}] p_0}
{4 [1 + e^{i \pi \mu}] \mu^3)}$
\\
$4$&$1$&
$i\frac{[-1 + e^{i \pi/\mu}] p_0}
{4 [1 + e^{i \pi/\mu}] \mu^4}$
\\
$4$&$0$&
$-i\frac{[-1 + e^{i \pi/\mu}] p_0}
{4 [1 + e^{i \pi/\mu}] \mu^4}$
\\
\hline
\hline
\end{tabular}
\caption{Quantum resonances up to fourth order in $\varepsilon$. The zeroth
  order resonance $k_{n,p}^{(0)}$ is here written modulo $2\pi/\mu^n$. Notice
  that the second order correction to $k_{2,p}$ is purely real. The same
  holds for the third and fourth order corrections to $k_{3,p}$. This
  suggests that the imaginary parts of $k_{n,p}$ are
  ${\mathcal O}(\varepsilon^{2n-1})$.}
\label{tab.knp}
\end{table}

In order to find the solution of 
${\mathsf  R}(k_{n,p})\phi_{n,p}=\phi_{n,p}$,
we will write again 
${\mathsf  R}(k_{n,p}) = {\mathsf
  R}^{(0)}(k^{(0)}_{n,p}) + \delta {\mathsf  R}(k_{n,p})$.
Upon expanding the eigenvectors $\phi_{n,p}$ in terms of the basis spanned
by the zeroth order eigenvectors $\phi^{(0)}_{n,0}$ and $\phi^{(0)}_{n,1}$,
$\phi_{n,p}=\sum_{m=1}^N\sum_{q=0,1}c_{n,p,m,q}\phi^{(0)}_{m,q}$,
we will make use of the property that $\phi^{(0)}_{m,q}$ is an eigenvector
of ${\mathsf R}^{(0)}(k^{(0)}_{n,p})$, 
${\mathsf R}^{(0)}(k^{(0)}_{n,p})\phi^{(0)}_{m,q}=
\Lambda_{n,p,m,q} \phi^{(0)}_{m,q}$,
with eigenvalue
$\Lambda_{n,p,m,q} = (-1)^q i \exp[(-1)^{p+1}i \pi \mu^{m-n}/2]$.
The first order correction to $k^{(0)}_{n,p}$ is the solution
$k^{(1)}_{n,p}$ of the equation
${\phi^{(0)}_{n,p}}^{\rm T}\delta {\mathsf R}(k^{(0)}_{n,p} + 
\varepsilon k^{(1)}_{n,p})\phi^{(0)}_{n,p} = 0$,
where $\delta{\mathsf R}$ must be expanded to first order in $\varepsilon$.
The only non-zero first order corrections have $n=1$, cf. Table \ref{tab.knp}.
The corresponding corrections to the zeroth order eigenvectors are given by
\begin{equation}
c_{n,p,m,q} = \frac{1}{\varepsilon}\frac{
{\phi^{(0)}_{m,q}}^{\rm T}\delta {\mathsf R}(k^{(0)}_{n,p} + \varepsilon k^{(1)}_{n,p})
\phi^{(0)}_{n,p}}
{1 - \Lambda_{n,p,m,q}}\ ,\quad\quad (m,q)\neq(n,p)\ ,
\end{equation}
which are different from zero for $(m=1,q=1-p)$ and $(m=2,q=0,1)$. One can
proceed to higher orders along these lines. The results for $k_{n,p}$ are 
presented in Table \ref{tab.knp} up to fourth order. We note that our
results suggest that the imaginary parts of the $k_{n,p}$ are ${\mathcal
  O}(\varepsilon^{2 n-1})$. Given that the distribution of the real parts
of the $k_{n,p}$ is rather uniform, this imply that the widths $\Gamma$
scale identically to the imaginary parts of the scattering
resonances. Hence Eq. (\ref{feq}) seems to hold for this example.

%%%%%%%%%%%%%%%%%%%%%%%%%%%%%%%%%%%%%%%%%%%%%%%%%%%%%%%%%%%%%%%%%%%%%%%%%%%%%
\section{conclusions}
\label{sec.conc}
%%%%%%%%%%%%%%%%%%%%%%%%%%%%%%%%%%%%%%%%%%%%%%%%%%%%%%%%%%%%%%%%%%%%%%%%%%%%%

We have presented a simple model of an open hierarchical graph for which an
analytical treatment of the algebraic decay of the survival probability is 
possible. The novelty of our approach lies on the successful application to 
the persistent hierarchical graph of a formalism originally developed in
the framework of fully 
chaotic systems, where the survival probability decays
exponentially. 

For the classical system, the computation of the survival probability
was done using the spectral decomposition of the evolution operator. 
We showed that the algebraic decay relies in an essential way on the
scaling properties of both the Pollicott-Ruelle resonances and their 
amplitudes. Using a pertubative approach we argued that the resonance
spectrum has an accumulation point at the value zero, which is
characterized by a
scaling property in terms of powers of the expansion parameter. The
structure of the corresponding eigenstates with respect to the length
scales of the system yields the scaling of the amplitudes. 

This result must be contrasted to the observation of algebraic decay in the 
self-similar Markov chains \cite{85HCM,ketzmerick}. Although the
exponents are identical, the hierarchical graph is a dynamical process
where randomness is involved only through the modelization of the
collisions with scatterers, as opposed to self-similar Markov chains where
transitions between states lack the spatial structure of our system. In 
the persistent hierarchical graph, the geometric role of the parameter
$\mu$ is very clear, whereas in
the self-similar Markov chains, $\mu$ represents an area which affects the
transition probabilities between states.

As already pointed out in the introduction, the parameters $\epsilon$ and
$\mu$ define a hierarchical dynamical trap \cite{98Z,99Z}, in the sense
that $\mu$ is the ratio between successive length scales and $\mu/\epsilon$
the ratio between the corresponding staying times. Our result
Eq. (\ref{deltaval}) is another instance of the relation of these
parameters to the transport properties of the system, in this case the
algebraic decay that characterizes the survival probability. It would be
interesting to know what relation does this bear to the transport exponent
of anomalous diffusion.

Other aspects of the properties of the classical persistent hierarchical
graph were studied through 
the application of the thermodynamic formalism. We computed the free
energy (or topological pressure) per unit time in terms of
the leading zero of a zeta function defined in analogy to discrete time
systems. Different asymptotic regimes were studied. In particular, the
topological entropy, which is the infinite temperature limit ($\beta\to 0$)
of the free energy, increases exponentially with the number
of bonds in the graph. In the limit of large number of bonds, the low 
temperature ($\beta\gg 1$)
free energy tends to zero exponentially with respect to the ratio
$\varepsilon/\mu<1$. Moreover these results suggest that the free energy 
undergoes a phase transition at $\beta=1$.

For the quantum system, we used methods similar to the classical
case and
conjectured that the widths of the quantum scattering resonances follow a
scaling law, in agreement with the numerically observed width distribution 
\cite{ketzmerick}. This argument was motivated by the computation of the
resonances to the first few orders in perturbation theory. The limitation
of this result, due to the complexity of the resolution of the quantum
problem, illustrates the gap that separates the understandings of the classical
and quantum approaches. The resolution of this question is open to future 
research by Bob Dorfman and others.

\begin{acknowledgments}
This paper is dedicated to our friend, mentor and colleague Bob Dorfman, on the
occasion of his 65th birthday. {\em Lechaim!}
The authors are grateful to Vered Rom-Kedar and Uzy Smilansky for their
helpful comments on the manuscript.
We thank the Israeli Council for Higher Education and the Feinberg 
postdoctoral fellowships program at the Weizmann Institute of Science for 
financial support.
F.~B. thanks the "Fundaci\'{o}n Andes" and their
program "inicio de carrera para jovenes cientificos" c-13760.
\end{acknowledgments}

%%%%%%%%%%%%%%%%%%%%%%%%%%%%%%%%%%%%%%%%%%%%%%%%%%%%%%%%%%%%%%%%%%%%%%%%%%%%%

%%%%%%%%%%%%%%%%%%%%%%%%%%%%%%%%%%%%%%%%%%%%%%%%%%%%%%%%%%%%%%%%%%%%%%%%%%%%%%
\end{document}